\documentclass[a4paper,prl,twocolumn, superscriptaddress]{revtex4}
\usepackage[utf8]{inputenc}
\usepackage{hyperref}
\usepackage{natbib}
\usepackage[separate-uncertainty]{siunitx}
\usepackage{amsfonts, amsmath, bm, graphicx}
\graphicspath{{figures/}}
\def\vec#1{{\bm{#1}}}

\def\t#1{{\mathrm{#1}}}

\begin{document}
\title{Spin transport-induced damping of coherent THz spin dynamics in iron}
\date{\today}

\author{L. Brandt}
\affiliation{Martin Luther University Halle-Wittenberg, Institute of Physics, Optics Group, Von-Danckelmann-Platz 3, 06120 Halle}
\author{R. Verba}
\affiliation{Institute of Magnetism, Kyiv 03142, Ukraine}
\author{N. Liebing}
\affiliation{Martin Luther University Halle-Wittenberg, Institute of Physics, Optics Group, Von-Danckelmann-Platz 3, 06120 Halle}
\author{M. Ribow}
\affiliation{Martin Luther University Halle-Wittenberg, Institute of Physics, Optics Group, Von-Danckelmann-Platz 3, 06120 Halle}
\author{I. Razdolski}
\affiliation{Faculty of Physics, University of Bialystok, 15-245 Bialystok, Poland}
\author{V. Tyberkevych}
\affiliation{Department of Physics, Oakland University, Rochester, MI 48309, USA}
\author{A. Slavin}
\affiliation{Department of Physics, Oakland University, Rochester, MI 48309, USA}
\author{G. Woltersdorf}
\affiliation{Martin Luther University Halle-Wittenberg, Institute of Physics, Optics Group, Von-Danckelmann-Platz 3, 06120 Halle}
\author{A. Melnikov}
\affiliation{Martin Luther University Halle-Wittenberg, Institute of Physics, Optics Group, Von-Danckelmann-Platz 3, 06120 Halle}

\begin{abstract}
We study the damping of perpendicular standing spin-waves (PSSWs) in ultrathin Fe films at frequencies up to 2.4 THz. The PSSWs are excited by optically generated ultrashort spin current pulses, and probed optically in the time domain. Analyzing the wavenumber and thickness dependence of the damping, we demonstrate that at sufficiently large wave vectors $k$ the damping is dominated by spin transport effects scaling with $k^4$ and limiting the frequency range of observable PSSWs. Although this contribution is known to originate in the spin diffusion, we argue that at moderate and large $k$ a more general description is necessary and develop a model where the 'transverse spin mean free path' is the a key parameter, and estimate it to be $\sim$0.5~nm.
\end{abstract}
\maketitle

Understanding dissipation mechanisms is an important problem of dynamical systems. Ferromagnetic materials exhibit a wide variety of dissipation processes contributing to the total decay rate of a spin waves (SW) depending on the material, static magnetization state, geometry, etc. \cite{Sparks_Book1964, Gurevich1996}, including 2-magnon scattering on imperfections  \cite{Sparks_PR1961, Arias_PRB1999, Hurben_JAP1998} and spin pumping into an adjacent nonmagnetic layers \cite{Tserkovnyak2002, Urban2001, Mizukami2001}. Originating in the magnon-phonon and magnon-electron interactions due to the spin orbit coupling, the most important contribution is often the intrinsic Gilbert damping $\alpha_G$ \cite{Gilbert_IEEE_TM2004}, resulting in the dissipation term $\vec T_d = \alpha_G/M_s \left(\vec M \times \partial \vec M/\partial t \right)$, where $\alpha_G$ is constant. In metallic ferromagnets, where the magnetic damping is dominated by the magnon-electron interactions, this phenomenological description is well supported by ab-initio calculations \cite{Gilmore_PRB2009,Kambersky_1972,Kunes2002,Gilmore2007, Seib2009, Steiauf2008}. Notably, the Gilbert model is local: the damping rate $\Gamma_n$ of a SW mode $n$ depends only on the mode frequency $\omega_n$ and precession ellipticity $\epsilon_n$, $\Gamma_n = \alpha_G \epsilon_n \omega_n $ \cite{Verba_PRB2018}, but not its wavenumber $k_n$. Within this framework, exchange-dominated magnons with $\omega_n \sim k_n^2$ exhibit the damping rate $\Gamma_n \sim k_n^2$. 

The phenomenological nonlocal, $k$-dependent dissipation \cite{Kambersky_1972} was shown to result in the $\Gamma_n \sim k_n^4$ dependence of the damping rate \cite{Bar_JETP1984, Bar_LTP2013}. Much later, transverse spin diffusion was proposed as a microscopic mechanism of the non-local spin dissipation \cite{Hankiewicz_PRB2008, Foros_PRB2008, Tserkovnyak_PRB2009, Wong_PRB2009, Zhang_PRL2009}, enhancing the effective Gilbert damping $\Delta\alpha_n \sim k_n^2$ \cite{Tserkovnyak_PRB2009, Wang_PRB2015}. The experimental verification of these predictions is scarce: in the microwave (GHz) frequency range (where most of the related research is performed) weak non-local damping effects are difficult to identify. Enhanced size-dependent damping was found in magnetic nanodots \cite{Nembach_PRL2013}, and $k_n^2$-dependence of the effective Gilbert damping parameter for perpendicular standing spin wave (PSSW) modes with $k_n<0.12\,\t{nm}^{-1}$ was confirmed \cite{Li_PRL2016}. Recently, non-monotonous $\alpha(k_n)$ dependence at $k_n\sim0.1\,\t{nm}^{-1}$ was reported in Co \cite{Lalieu2019}. 

In this letter, we study PSSWs in Fe and demonstrate that an exchange-driven spin transport mechanism plays the dominant role in the magnetization damping at large wavenumbers. Owing to the rapid increase of its efficiency with increasing wavenumber, at $k\gtrsim$2~nm$^{-1}$ (and frequencies beyond 2.4~$\t{THz}$) the damping rate exceeds the SW eigenfrequency, and the PSSWs become overdamped. However, the extrapolation of this trend to even larger wavenumbers approaching $\pi/a$ contradicts well-established measurements that are based e.g. on neutron scattering  \cite{Shirane1968,Mook1973,Prokop2009,Zakeri_PRL2010,Zhang2012}. By describing the transverse spin transport in a ballistic model we show that spin-transport induced damping saturated at very large wave vectors and resolve this apparent contradiction.

\begin{figure}[htbp]
\includegraphics[width=1\linewidth]{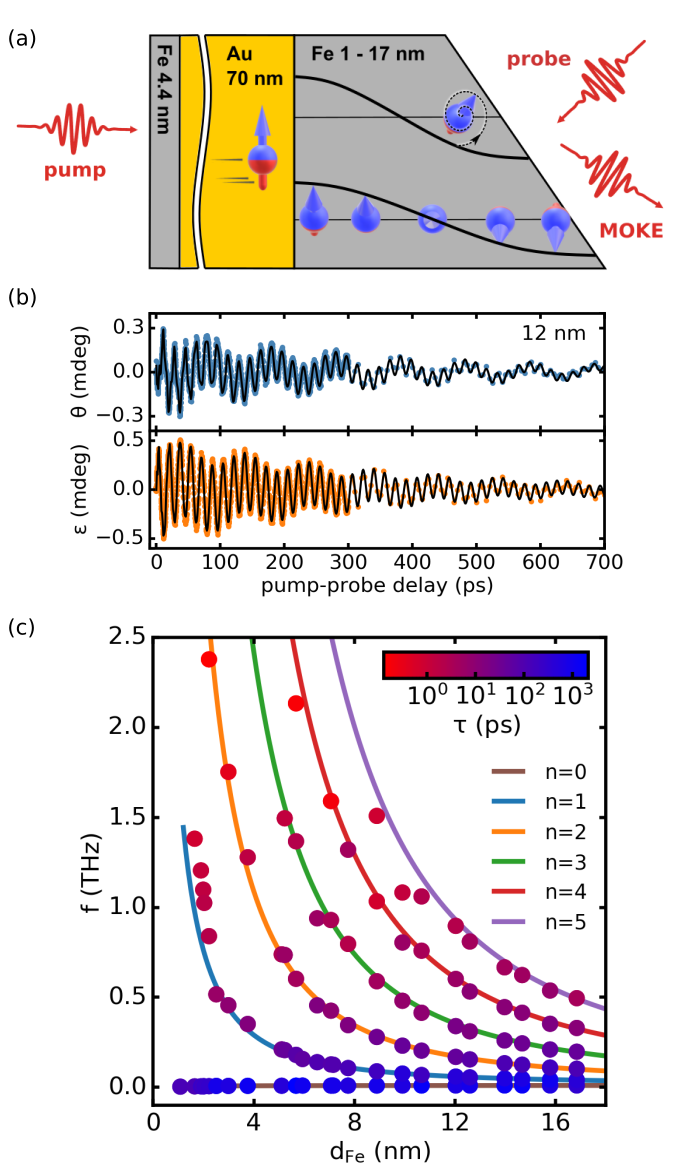}
	\caption{(a) Sample layout and measurement geometry. (b) Exemplary time-resolved MOKE signals (rotation $\theta$ and ellipticity $\varepsilon$) at 12-nm-thick Fe collector including the fit with damped cosines. (c) FMR and PSSW frequencies (y-axis) and lifetimes (color-code) obtained by fits to the time-resolved MOKE traces (as shown in (b)). The solid lines are a theoretical fit \cite{Brandt2021}. }
	\label{fig:raw}
\end{figure}

For our experiments epitaxial Fe/Au/Fe(001) multilayers (Fig. \ref{fig:raw}a), are prepared \cite{Brandt2021}. The THz-frequency spin dynamics is measured employing the optical back pump-front probe technique used in \cite{Melnikov2011,Alekhin2017,Razdolski2017}. Spin dynamics in the Fe wedge (collector) is excited by an ultrashort spin current pulse generated in the Fe emitter by a femtosecond laser stimulus. Spin transfer torque (STT) driven PSSWs \cite{Razdolski2017} are detected in the time domain using the magneto-optical Kerr effect (MOKE). Maintaining the fs time resolution, the 70 nm-thick Au spacer precludes the concomitant optical excitation of the collector \cite{Melnikov2021}. 
This experimental approach allows for systematic measurements of the frequencies and lifetimes of the PSSW modes vs. the collector thickness using a single sample.

In Fig. \ref{fig:raw}(b) the time-resolved polar MOKE rotation and ellipticity are shown. The periodic signals with multiple frequencies indicate the excitation of several PSSW eigenmodes. By fitting these curves with a sum of damped cosine functions a set of lifetimes $\tau_n=\Gamma_n^{-1}$ and frequencies $f_n$ for each PSSW eigenmode with index $n$ is obtained. The observed modes can be identified by the PSSW dispersion accounting for the magnetic anisotropy and the exchange stiffness at different Fe thicknesses \cite{Brandt2021}. The PSSW wavenumbers are well described by $k_n=n\pi/d_{\mathrm{Fe}}$ implying that there are no magnetically dead layers. Up to 6 spin wave modes are observed (the ferromagnetic resonance (FMR) mode and 5 PSSW modes), depending on the collector thickness $d$. The maximum observed PSSW frequency is about $f_{\mathrm{max}} =$ \SI{2.4} {\tera \hertz} (Fig. \ref{fig:raw}(c)), which corresponds to the shortest observed wavelength $\lambda_{\mathrm{min}} =$ \SI{2.22} {\nano \metre}.

\begin{figure*}[htbp]
    \includegraphics[width=1\linewidth]{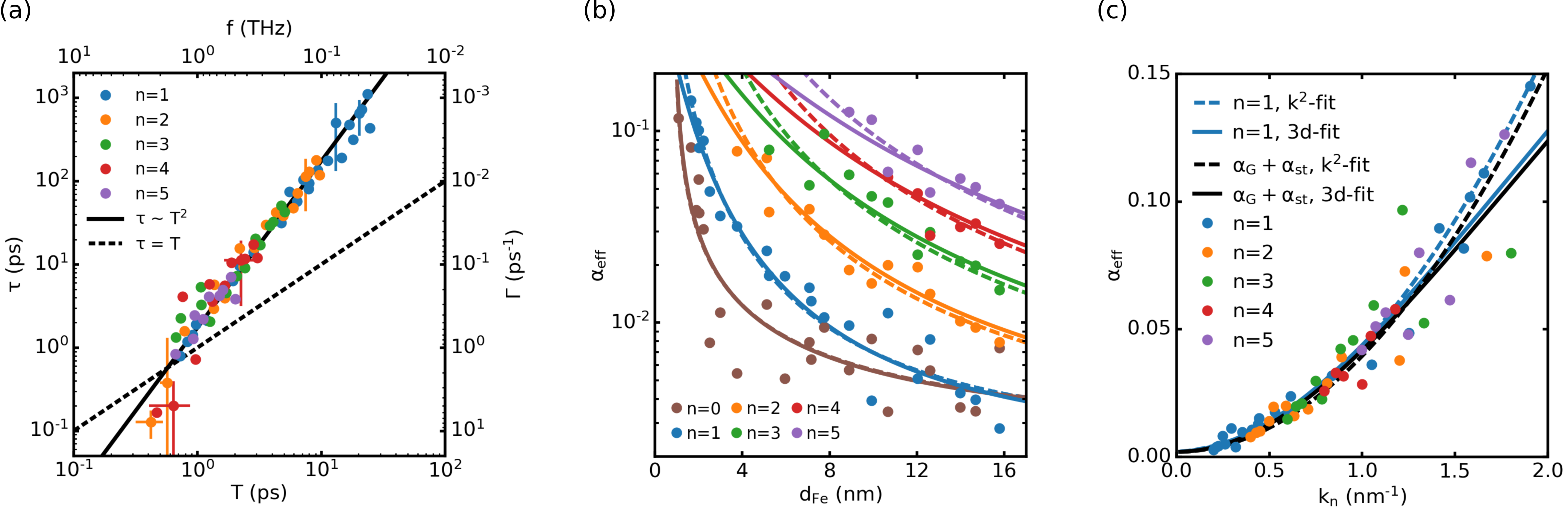}
    \caption{(a) Double logarithmic plot of measured lifetimes $\tau$ (damping rate $\Gamma$) of PSSW modes vs.~the mode oscillation period $T$ (frequency $f$); solid line represents the quadratic fit $\tau \sim T^2$ and dashed line represents the dependence $\tau = T$.  Effective damping as a function of the Fe thickness (b) and PSSW wave vector (c); points are the experimental data, dashed and solid lines represent the fit with Eq.~(\ref{eq:alpha}) and 3d model (see text), respectively.}
	\label{fig:damping}
\end{figure*}

The dependence of the SW lifetime on the oscillation period $T_n=1/f_n$ is shown in Fig.~\ref{fig:damping}(a). It is evident that the data are well fitted by a $\tau_n \sim T_n^2$ dependence (the solid line) corresponding to $\Gamma_n \sim k_n^4$ when $f_n \sim k_n^2$. We find that at $f \approx 1.8\,\t{THz}$ ($T \approx 0.56\,\t{ps}$, see the intersection with the $\tau_n = T_n$ dashed line in Fig.~\ref{fig:damping}(a)) the SW lifetime becomes smaller than the oscillation period, indicating the overdamped regime. Above this critical frequency the PSSWs are hardly detectable, and only very few data points with large uncertainties are obtained between 1.8 and 2.4~THz but discarded below.

We further discuss an effective Gilbert damping parameter $\alpha_{\t{eff},n} = \Gamma_n/(\omega_n \epsilon_n) = (2\pi \tau_n f_n \epsilon_n)^{-1}$, which is convenient for comparing the relaxation rates between the modes. The ellipticity-related coefficient $\epsilon_n$ was calculated 
from the SW dispersion \cite{Brandt2021, Verba_PRB2018}. In the range of interest (above 100 GHz) $\epsilon_n\approx 1$ while for the FMR mode $\epsilon_0\approx 3$. In Fig.~\ref{fig:damping}(b) it is seen that the effective damping parameter $\alpha_{\t{eff},n}$ strongly increases with the mode number $n$ and the inverse film thickness. 
To analyze these data, we employ the following phenomenological non-local damping model:
\begin{equation}
\alpha_{\mathrm{eff},n}= \alpha_G + \frac{\Gamma_{\mathrm{nu}}}{\omega_n \epsilon_n} + \frac{c_n G_{\mathrm{nu}}}{\omega_n \epsilon_n d} +  \frac{c_n \eta_{\mathrm{s}}}{d} + \eta_{\perp}k_n^2 \ .
\label{eq:alpha}
\end{equation}
The first term in this expression represents the intrinsic Gilbert damping, the next two terms describe the non-uniform resonance line broadening with volume contribution $\Gamma_\t{nu}$ (due to volume defects) and interface contribution $G_\t{nu}/d$ (due to interface imperfections). The impact of the latter naturally scales inversely with the film thickness, as well as the fourth term $\sim \eta_s$ describing spin-pumping (spin-current emission) into the adjacent Au layer. The mode-profile-dependent coefficients are $c_n=1$ for $n=0$ and $c_n=2$ for the higher modes \cite{Verba_PRB2018}. The last term represents the $k_n^2$-dependent exchange-driven contribution $\alpha_{st}$ which we attribute to spin transport.

\begin{figure}
	\includegraphics[width=1\columnwidth]{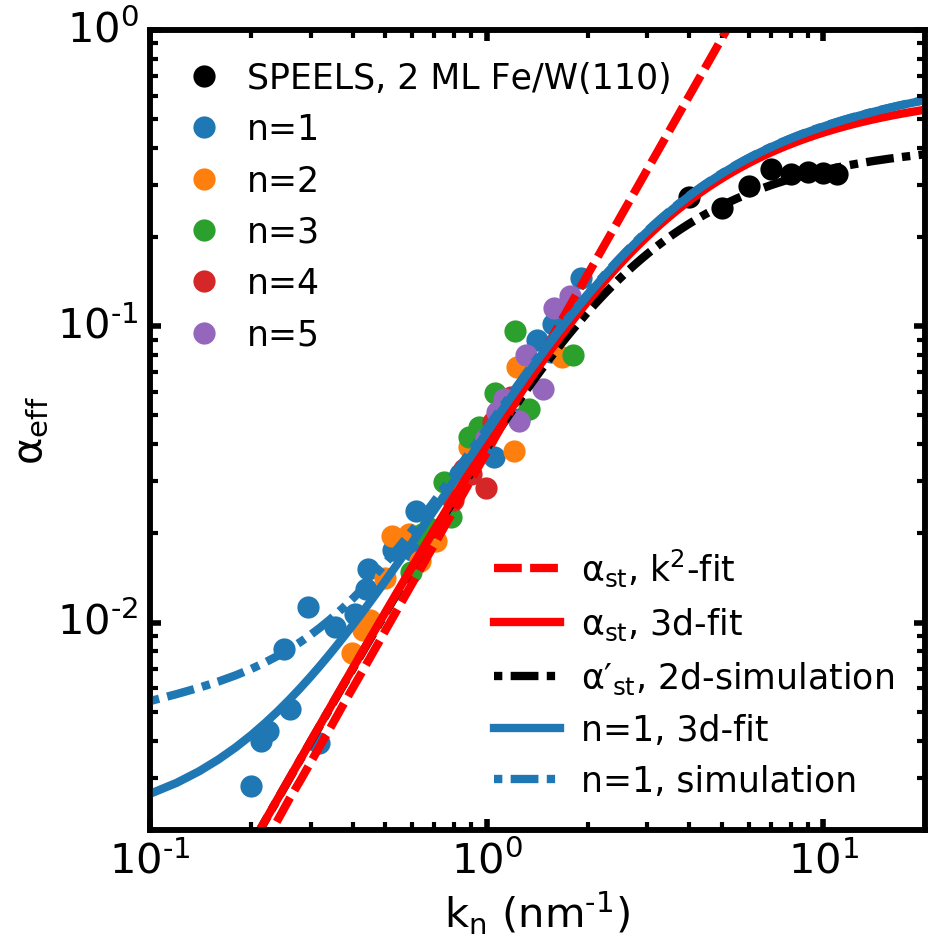}
	\caption{Effective damping of PSSW modes as a function of the wave vector along with SPEELS data from Ref.~\cite{Zhang2012}; curves represent model fits (see legend and text).}
	\label{fig:MOKEandSPEELS}
\end{figure}

Eq.~(\ref{eq:alpha}) allows to separate contributions to the effective damping parameter $\alpha_{\mathrm{eff},n}$ owing to their different dependencies on the collector thickness \cite{note1}. The joint fit (Fig.~\ref{fig:damping}(b,c), dashed curves) of all 6 measured SW modes (i) corroborates our understanding of the damping, and (ii) unambigously  demonstrates the dominant role of the spin transport contribution $\alpha_{st}=\eta_\bot k_n^2$. This term arises from the transport of transverse spin component along the SW $k$-vector and follows the rising curvature of spatial mode profile with $\eta_\bot = (38\pm 3) \cdot 10^{-3}\,\t{nm}^{2}$. The spin pumping term $\alpha_{sp}=\eta_s c_n/d\propto k_n$ scales as $n^{-1}$ and is responsible for the small deviation from the parabola (compare blue and black curves in Fig.~\ref{fig:damping}(c)). It never exceeds 5$\%$ of the total damping and in the following we use the value $\eta_{\mathrm{s}}=3 \cdot 10^{-3}\, \t{nm}$ determined for Fe/Au interfaces in previous studies \cite{Urban2001,Heinrich2003}. 

Our results show that \emph{local} damping terms in Eq.~(\ref{eq:alpha}) provide only a minor contribution to $\alpha_\mathrm{eff}$. The Gilbert damping $\alpha_G= (2.0 \pm 0.7) \cdot 10^{-3}$ is close to that reported earlier \cite{Urban2001,Heinrich2003} and matches the recent results \cite{Wu2021}. The contribution of sample inhgomogeneity  vanishes and we find $\Gamma_\t{nu}=(0 \pm 0.2) \cdot 10^{-3}\,\mathrm{ps}^{-1}$ confirming the the high epitaxial quality of the sample. Similarly, the interface defect $G_\t{nu}$ contribution also vanished for all $n\neq 0$ modes. It is only significant for the $n = 0$ mode, where it dominates the relaxation rate with $G_\t{nu}=(6.0 \pm 0.7) \cdot 10^{-3}\, \mathrm{nm}\, \mathrm{ps}^{-1}$. 
The absence of this term for for higher modes is illustrated in Fig.~\ref{fig:MOKEandSPEELS}  where the dash-dotted blue curve is simulated by adding this contribution to the fit results (solid blue curve) for the 1-st PSSW mode. The characteristic increase at small $k_n$ is not reproduced by the experimental data indicating the absence of interface two-magnon scattering for PSSWs with $n\geq1$.  

Despite the unambiguous dominance of the $k^2$ spin transport contribution at $k<2$~nm$^{-1}$, we argue that a saturation of this term is expected at larger $k$ \cite{note3}. Otherwise, as pointed out by Baryakhtar \cite{Bar_LTP2013}, SW at even larger $k$ would experimentally not be detectable by the neutron scattering or SPEELS  \cite{Shirane1968,Mook1973,Zakeri_PRL2010,Prokop2009,Zhang2012}. This is illustrated in Fig.~\ref{fig:MOKEandSPEELS} where the effective damping calculated as a ratio of half width at half maximum to the peak energy from the SPEELS data obtained on a 2-ML-thick Fe/W(110) \cite{Zhang2012} is shown together with our MOKE results. The parabolic extrapolation overestimates the damping by one order of magnitude for $k\approx 10$~nm$^{-1}$. This saturation at $k_s \gtrsim 6$~nm$^{-1}$ hints at a characteristic length scale $\lambda_s=\pi/k_s\lesssim 1$~nm, much smaller than the spin diffusion length $\lambda_\t{sd}^\t{Fe}=7$~nm \cite{Ko2020}. As such, in order to describe MOKE and SPEELS results on equal footing, spin transport needs to be analyzed beyond the diffusion model resulting in a non-local dissipative term $-\eta_\bot \vec m \times \nabla^2 \partial \vec m/\partial t$, where $\vec m = \vec M/M_s$ is the reduced magnetization unit vector \cite{Tserkovnyak_PRB2009,Li_PRL2016}.

To make a step towards such analysis, we keep the time-dependent part but revise the spatial non-locality description considering the transverse spin transport in itinerant ferromagnets based on the \emph{s-d} model. The transport is provided by quasi-free $s$-electrons traveling with the Fermi velocity $v_F\sim1$~nm/fs in a 'spin jellium' formed by quasi-localized $d$-moments precessing at a frequency $\omega$ owing to the exchange interaction. Exchanging the transverse spin component with $s$-electrons, $d$-moments can change the angle (relaxation) or the phase (dephasing) of precession obeying the angular momentum conservation. Owing to small precession angles usually observed in experiments, the longitudinal spin component remains almost constant. For $s$-electrons this interaction results mainly in a phase shift between the majority and minority components of their wave function \cite{note4}. Therefore, neither the energy nor momentum of the $s$-electrons change and such processes become very probable, so that even a sub-fs time scale can be expected owing to the broad, $\sim 10$~eV $s$-band and strong \emph{s-d} exchange. The corresponding sub-nm length scale, 'transverse spin mean free path' $\lambda_s$ is then much shorter than the $s$-electron inelastic mean free path at the Fermi level $\lambda_e \gtrsim 10$~nm \cite{Zhukov2006,note5}. Thus, we consider $s$-electrons moving ballistically with isotropic velocity distribution and dragging the transverse  component of spin momentum along their trajectory. 
In a ferromagnet with spatially non-uniform dynamic magnetization, such spin transfer can be viewed as ferromagnetic spin pumping in the continuum limit \cite{Tserkovnyak_PRB2009} resulting in the non-local damping considered here.

For a quantitative analysis, we consider the balance of transverse spin momentum dragged by \emph{s}-electrons away from and into a given volume around $\vec r_0$. We assume that at $\vec r_0$ the transverse magnetization component $\vec M_\bot \cos{k_n z_0}$ decays at a rate $\xi \omega_n$ since it is redistributed into the surrounding with a spin density $\propto \exp{(-|\vec r - \vec r_0|/\lambda_s)}$. Integration over the volume results in the following expression for the spin transport contribution to the damping (the last term in Eq.~(\ref{eq:alpha})): $\alpha_{st} = \xi [1-\arctan{(\lambda_s k_n)}/(\lambda_s k_n)]$. By setting $\xi=3\eta_\bot/\lambda_s^2$ we recover the initial form $\alpha_{st} \approx \eta_\bot k_n^2$ for small $k$. Fitting the experimental data to this expanded spin transport model (solid curves in Fig.~\ref{fig:damping}(b,c)) yields $\lambda_s = 0.46 \pm 0.16$~nm. This value is indeed in the expected range as discussed above supporting our considerations. Remarkably, we find a value for the transverse spin mean free path, which is in agreement with the STT penetration depth calculated \cite{Stiles_PRB02} at Fe interfaces. One expects $\lambda_{STT} \lesssim \lambda_s$, in agreement with the earlier estimated $ \lambda_{\mathrm{STT}} \lesssim$ \SI{0.56}{\nano\metre} $\approx$ 4~ML \cite{Brandt2021}.  The obtained value of $\eta_\bot = (45\pm 6) \cdot 10^{-3}\,\t{nm}^{2}$ (slighlty larger cmpared to the parabolic fit) is about a factor of two smaller compared to previous FMR results obtained for NiFe, Co, and CoFeB at small k-vectors \cite{Li_PRL2016}. The corresponding transverse conductivity is  $\sigma_\bot = \eta_\bot M_s/\gamma = 4.4\cdot10^{-25}\,\t{J s}/\t{m}$ \cite{Verba_PRB2018}, where $\gamma=1.76\cdot10^{11}\t{(s T)}^{-1}$ is the gyromagnetic ratio. 

In Fig.~\ref{fig:MOKEandSPEELS}, $\alpha_{st}$ obtained using  a 3d-integration is shown by the red solid curve. At $k>0.2$~nm$^{-1}$ $\alpha_{st}$ already exceeds $\alpha_G=0.002$ (corresponding to the bottom of the graph). The deviations from the parabolic behaviour (red dashed line) become significant for $k > 1$~nm$^{-1}$, when the distance between the adjacent SW antinodes $\lambda/2$ becomes comparable to $\lambda_s$. At much larger $k$ vectors, when $\lambda \ll \lambda_s$, the transverse spin density delivered to a given volume from the surrounding averages to zero, and therefore $\alpha_{st}$ is expected to saturate at $\xi$. Qualitatively this result is in agreement with the experimental SPEELS data. However, the red curve in Fig.~\ref{fig:MOKEandSPEELS} still overestimates the damping observed by SPEELS at $k>5$~nm$^{-1}$ by about 30$\%$. 
We note that in contrast to MOKE experiments the sample used for the SPEELS experiment is an extremely thin Fe film of only 2 ML thickness ($d\approx0.15$~nm~$ < \lambda_s$).
Accordingly, the 2d character of the spin transport requires an integration only in the plane (not 3d) resulting in $\alpha^{\prime}_{st} = \xi^{\prime} [1-1/\sqrt{1+\lambda_s^{\prime 2} k^2}]$, where $\xi^{\prime}=2\eta^{\prime}_\bot/\lambda_s^{\prime 2}$. Strictly speaking, both $\lambda_s^{\prime}$ and $\eta^{\prime}_\bot$ can differ from $\lambda_s$ and $\eta_\bot$ in the 3d system. However, a simulation with $\lambda_s^{\prime}=\lambda_s$ and $\eta^{\prime}_\bot=\eta_\bot$ (black dashed curve) already describes the SPEELS data very well.
Further experimental observations of high-$k$ SWs and their damping rates in various ferromagnets are thus highly desirable for the verification of this model, paving a way towards understanding spin transport at the nanoscale.

In summary, we have demonstrated that the spin transport-driven contribution dominates the damping of THz-frequency PSSWs in thin Fe films. Our results show that at moderate wavenumbers $k \lesssim 1\,\t{nm}^{-1}$ (up to about 0.25~THz frequency), this transport can be conventionally treated as transverse spin diffusion, resulting in the $k_n^2$-dependence of the effective damping parameter $\alpha_{\t{eff},n}$. For the effective spin-wave damping rate this corresponds to the $\Gamma_n \sim k_n^4$ dependence. At larger $k$ (and shorter spatial scales), the super-diffusive or ballistic character of the spin transport needs to be taken into account. The rapid increase of $\alpha_{\t{eff},n}$ with $k$ causes the damping rate in Fe to exceed the eigenfrequency of the studied PSSWs above 1.8~THz, preventing measuremts at higher frequencies using MOKE. To analyze our results on equal footing with those obtained by SPEELS at much larger k-values (and frequencies) we introduce a 'ballistic' model of transverse spin transport.  This model well accounts for the experimental observations in the entire range of $k$-vectors and allows to identify a 'transverse spin mean free path' $\lambda_s\approx0.5$~nm in Fe. Our findings advance the understanding of spin transport at the nanoscale and enable further investigations of coherent STT excitation of spin waves. In particular, the next step towards nanoscale THz spintronics entails studying SWs in FM and AFM materials at even larger wavenumbers and analyzing the crossover between the diffusive and ballistic spin transport regimes.

This work was supported by the German research foundation (DFG) through CRC/TRR 227 (project B01), the National Science Center Poland (Grant No. DEC-2019/35/B/ST3/00853), U.S. National Science Foundation (Grant \#EFMA-1641989), by the Air Force Office of Scientific Research under the MURI grant \#FA9550-19-1-0307,  by the DARPA TWEED grant \#DARPA-PA-19-04-05-FP-001, by the Oakland University Foundation, and by the Ministry of Education and Sciences of Ukraine (project \#0121U110090).


\begin{thebibliography}{46}
\bibitem[{not({\natexlab{b}})}]{note3}
\bibinfo{note}{Fig.~\ref{fig:damping}(c) already makes an impression that the
  dashed curves overestimate the damping at $k>1$~nm$^{-1}$.}
\bibitem[{not({\natexlab{c}})}]{note4}
\bibinfo{note}{Here we suggest that in most of such events the $s$-spin is
  already tilted after previous 'spin collisions', which minimizes the change
  of amplitude ratio of the majority and minority wave-function components
  (determining the longitudinal spin projection).}
\end{thebibliography}
\end{document}